\documentstyle[referee]{laa}

\begin{document}
\thesaurus{07  
	   (07.09.1; 
	    07.13.1;  
	   )}

\title{Electromagnetic Radiation and Equation of Motion for Really Shaped
Particle -- New Covariant Formulation}
\author{J.~Kla\v{c}ka}
\institute{Institute of Astronomy,
   Faculty for Mathematics and Physics, Comenius University \\
   Mlynsk\'{a} dolina, 842~48 Bratislava, Slovak Republic}
\date{}
\maketitle

\begin{abstract}
Relativistically covariant form of equation of motion for real particle
(body) under the action of electromagnetic radiation is derived.
Equation of motion in the proper frame of the particle uses the radiation
pressure cross section 3 $\times$ 3 matrix.

Obtained covariant equation of motion is compared with another covariant
equation of motion which was presented more than one year ago.


\end{abstract}

\section{Introduction}
Covariant form of equation of motion for a particle is required if one wants
to be sure that the motion of the particle is correctly described.
Kla\v{c}ka (2000a) has derived covariant form of equation of motion for
really shaped dust particle under the action of electromagnetic radiation.
Derivation to the first order in $v/c$ (higher orders are neglected;
$\vec{v}$ is velocity of the particle, $c$ is the speed of light) was presented
in Kla\v{c}ka (2000b) and application to large bodies (asteroids) can be found
in Kla\v{c}ka ( 2000c). Later on, a little different derivation to the
first order in $v/c$ together with application to real system
in the Solar System was presented (Kla\v{c}ka and Kocifaj 2001).

The aim of this paper is to present a new covariant form of the
equation of motion for the discussed problem. It will be of a form
similar to that for Lorentz force -- product of a tensor of the second rank
with a four-vector. The crucial question which of the two equations
of motion (that presented in Kla\v{c}ka (2000a) and that presented in this
paper) is correct, will be discussed in detail.

\section{Proper reference frame of the particle -- stationary particle}

Primed quantities will denote quantities measured in the
proper reference frame of the particle.

Equation of motion of the particle in its proper frame of reference
is taken in the form
\begin{eqnarray}\label{1}
\frac{d~ E'}{d~ \tau} &=& 0 ~,
\nonumber \\
\frac{d~ \vec{p'}}{d~ \tau} &=& \frac{1}{c} ~ S'~ \left ( C ' ~
				\vec{S} ' \right ) ~,
\end{eqnarray}
where $E'$ is particle's energy, $\vec{p'}$ its momentum, $\tau$ is proper
time, $c$ is the speed of light, $S'$ is the flux density of radiation energy
(energy flow through unit area perpendicular to the ray per unit time),
$C'$ is radiation pressure cross section 3 $\times$ 3 matrix,
unit vector $\vec{S} '$ is directed along the path of the incident
radiation (it is supposed that beam of photons propagate
in parallel lines) and its orientation corresponds to the orientation
of light propagation.

Eqs. (1) describe equation of motion of the particle in the proper
frame of reference due to its interaction with electromagnetic radiation.
(It is supposed that the energy $E'$ of the particle is unchanged:
the energy of the incoming radiation equals
to the energy of the outgoing radiation, per unit time.)

\section{Covariant equation of motion}

We want to derive equation of motion for the particle in the
frame of reference in which particle moves with actual velocity $\vec{v}$.
We will use the fact that we know
this equation in the proper frame of reference -- see Eqs. (1).

Inspiration comes from the fact that space components of four-momentum
are written as a product with unit vector $\vec{S} '$. We know that
this unit vector can be generalized to four-vector (see, e. g., Kla\v{c}ka 2000a)
\begin{eqnarray}\label{2}
b^{\mu} &=& \left ( 1 ~/~ w, \vec{S} ~/~w \right )
\nonumber \\
w &\equiv& \gamma ~ ( 1 ~-~ \vec{v} \cdot \vec{S} / c ) ~, ~~
	   \gamma \equiv 1 ~/~ \sqrt{1 ~-~ ( \vec{v} ~/~ c )^{2}} ~.
\end{eqnarray}
Moreover, we know that $w^{2} ~S ~/~ c$ is a scalar quantity -- invariant
of the Lorentz transformation (see, e. g., Kla\v{c}ka 2000a).

Idea is to write covariant equation of motion in the form
\begin{equation}\label{3}
\frac{d~ p^{\mu}}{d~ \tau} = \frac{w^{2}~S}{c} ~ G^{\mu~\nu} ~
			     b_{\nu} ~.
\end{equation}
The only problem is to find components of the tensor of the second rank
$G^{\mu~\nu}$.

In order to find $G^{\mu~\nu}$, we will proceed in two steps. At first, we
will rewrite Eq. (3) in the proper frame of the reference of the particle.
Comparison with Eqs. (1) will yield components of $G^{' ~\mu~\nu}$.
The second step is transformation from $G^{' ~\mu~\nu}$ to $G^{\mu~\nu}$.

As for the proper frame of reference, Eq. (3) yields
\begin{eqnarray}\label{4}
\frac{d~ E'}{d~ \tau} &=& \frac{w^{2}~S}{c} ~\left \{ G^{'~0~0} ~-~
\sum_{j=1}^{3} G^{'~0~j} ~ \left ( \vec{S} ' \right ) _{j} \right \} ~,
\nonumber \\
\frac{d~ \left ( \vec{p'} \right ) _{k}}{d~ \tau} &=&  \frac{w^{2}~S}{c} ~
\left \{ G^{'~k~0} ~-~
\sum_{j=1}^{3} G^{'~k~j} ~ \left ( \vec{S} ' \right ) _{j} \right \} ~,
\end{eqnarray}
where the term $1/w'$ in the brackets is omitted due to a simple fact that
it equals 1 in the proper frame of reference. Comparison with Eqs. (1) yields
\begin{eqnarray}\label{5}
G^{'~0~0} = G^{'~0~j} = G^{'~j~0} = 0 ~, ~~ j = 1, 2, 3 ~,
\nonumber \\
G^{'~k~j} = -~ C '_{k~j} ~, ~~j = 1, 2, 3 ~, ~~k = 1, 2, 3 ~.
\end{eqnarray}

In order to find $G^{\mu~\nu}$, we have to use Lorentz transformation
\begin{equation}\label{6}
G^{\mu ~\nu} = \Lambda_{\alpha}^{~~\mu} ~ \Lambda_{\beta}^{~~\nu} ~
	       G^{'~\alpha~\beta} ~,
\end{equation}
where summation over repeated indices is supposed (and also in all the following
equations) -- 0, 1, 2, 3 for greek letters and 1, 2, 3 for latin letters --
and
\begin{eqnarray}\label{7}
\Lambda_{\alpha}^{~~\beta} &=& \eta _{\alpha ~ \rho} ~ \eta ^{\beta ~ \gamma} ~
			      \Lambda^{\rho}_{~~\gamma} ~,
\nonumber \\
\eta _{\alpha ~ \beta} &=& diag ( +~1, -~1, -~1, -~1 ) ~.
\end{eqnarray}
Generalized special Lorentz transformation yields
\begin{eqnarray}\label{8}
\Lambda_{0}^{~~0} &=& \gamma
\nonumber \\
\Lambda_{0}^{~~i} &=& \Lambda_{i}^{~~0} = \gamma ~( \vec{v} / c )_{i}
		      \equiv \gamma ~ ( \vec{\beta} )_{i} ~, ~~ i = 1, 2, 3 ~,
\nonumber \\
\Lambda_{i}^{~~j} &=& \delta_{i~j} ~+~ ( \gamma ~-~ 1 ) ~( \vec{\beta} )_{i}
		      ~ ( \vec{\beta} )_{j} ~/~ \vec{\beta}^{2}   ~,
		     ~~ i = 1, 2, 3 ~,~~ j = 1, 2, 3 ~.
\end{eqnarray}

Inserting Eqs. (5) and (8) into Eq. (6), one can find all components
of the tensor $G^{\mu~\nu}$:
\begin{eqnarray}\label{9}
G^{0~0} &=& -~ \gamma^{2} ~( \vec{\beta} )_{i} ~( \vec{\beta} )_{j} ~ C'_{i~j}
     \equiv -~ \gamma^{2} ~\vec{\beta} ^{T} ~(C' ~ \vec{\beta} ) ~,
\nonumber \\
G^{0~i} &=& -~ \gamma ~( \vec{\beta} )_{j} ~ C'_{j~k} ~ \left \{
	    \delta_{k~i} ~+~ ( \gamma ~-~ 1 ) ~( \vec{\beta} )_{k} ~
	    ( \vec{\beta} )_{i} ~/~ \vec{\beta}^{2} \right \}
\nonumber \\
     &\equiv& -~ \gamma ~ ( \vec{\beta} )_{j} ~C'_{j~i} ~-~ \gamma ~
	      ( \gamma ~-~ 1 ) ~\left [ ( \vec{\beta} )^{T} ~
	      ( C' ~\vec{\beta} ) \right ] ~( \vec{\beta} )_{i}
	      ~/~ \vec{\beta}^{2}  ~,
\nonumber \\
G^{i~0} &=&
	   -~ \gamma ~( \vec{\beta} )_{k} ~ \left \{  C'_{i~k} ~+~
	   ( \gamma ~-~ 1 ) ~( \vec{\beta} )_{i} ~
	    ( \vec{\beta} )_{j} ~C'_{j~k} ~/~ \vec{\beta}^{2} \right \}
\nonumber \\
     &\equiv& -~ \gamma ~ ( C' ~\vec{\beta} )_{i} ~-~ \gamma ~
	      ( \gamma ~-~ 1 ) ~\left [ ( \vec{\beta} )^{T} ~
	      ( C' ~\vec{\beta} ) \right ] ~( \vec{\beta} )_{i}
	      ~/~ \vec{\beta}^{2}  ~,
\nonumber \\
G^{i~j} &=& -~ C'_{k~l} ~
	    [ \delta_{i~k} ~+~ ( \gamma ~-~ 1 ) ~( \vec{\beta} )_{i} ~
	    ( \vec{\beta} )_{k} ~/~ \vec{\beta}^{2} ] ~
	    [ \delta_{l~j} ~+~ ( \gamma ~-~ 1 ) ~( \vec{\beta} )_{l} ~
	    ( \vec{\beta} )_{j} ~/~ \vec{\beta}^{2} ]
\nonumber \\
     &\equiv& -~ C'_{i~j} ~-~ ( \gamma ~-~ 1 ) ~( \vec{\beta} )_{j} ~
	      ( C' ~\vec{\beta} )_{i} ~/~ \vec{\beta}^{2}  ~-~
	      ( \gamma ~-~ 1 ) ~( \vec{\beta} )_{k} ~ C'_{k~j}~
	      ( \vec{\beta} )_{i} ~/~ \vec{\beta}^{2} ~-~
\nonumber \\
	& &   [ ( \gamma ~-~ 1 ) ~/~ \vec{\beta}^{2} ]^{2} ~
	      \left [ ( \vec{\beta} )^{T} ~( C' ~\vec{\beta} ) \right ] ~
	      ( \vec{\beta} )_{i} ~ \vec{\beta}_{j}  ~.
\end{eqnarray}
It is worthwhile to mention that $C'_{i~j} = A' ~ \delta _{i~j}$ leads to
symmetric tensor $G^{\mu~\nu}$.

Instead of making algebra in putting Eqs. (9) into Eq. (3),  we will
make more important procedure. We want to prove that Eq. (3) yields
result consistent with more transparent Eq. (28) in Kla\v{c}ka (2000a).

\section{Consistency of the known covariant formulations}

We have obtained equation of motion in the form of Eq. (3) and we have determined
the components of quantities on its right-hand side. Another form of
covariant equation was presented in Eq. (28) in Kla\v{c}ka (2000a).
Are these equations consistent?

Kla\v{c}ka (2000a) presents covariant equation of motion in the form
corresponding to
\begin{equation}\label{10}
\left ( \frac{d~ p^{\mu}}{d~ \tau} \right ) _{I} = \frac{w^{2}~S~A'}{c} ~
	   \sum_{j=1}^{3} Q'_{j} \left (
	   b_{j}^{\mu} ~-~ \beta^{\mu} \right ) ~;
\end{equation}
it is supposed that $b_{1}^{\mu} = b^{\mu}$. This paper has discussed
equation of motion of the form
\begin{equation}\label{11}
\left ( \frac{d~ p^{\mu}}{d~ \tau} \right ) _{II} = \frac{w^{2}~S}{c} ~
	  G^{\mu~\nu} ~ b_{\nu} ~.
\end{equation}
We want to show that Eqs. (10) and (11) are equivalent, i. e., that
$\left ( d~ p^{\mu} ~/~ d~ \tau \right ) _{I}$ $=$
$\left ( d~ p^{\mu} ~/~ d~ \tau \right ) _{II}$.

Multiplication of Eq. (10) by four-vector $b_{k~\mu}$ (and summation over
$\mu$) yields
\begin{equation}\label{12}
\left ( \frac{d~ p^{\mu}}{d~ \tau} \right ) _{I} ~b_{k~ \mu}  =
	\frac{w^{2}~S~A'}{c} ~\sum_{j=1}^{3} Q'_{j} \left (
	b_{j}^{\mu}~b_{k~ \mu}	~-~ \beta^{\mu}~ b_{k~ \mu}  \right )
				~~, ~k = 1, 2, 3 ~.
\end{equation}
It can be easily verified, on the basis of Eq. (2) and analogous equations
for $b_{2}^{\mu}$ and $b_{3}^{\mu}$ (see Eqs. (27) in Kla\v{c}ka 2000a), that
\begin{equation}\label{13}
\beta^{\mu} ~b_{k~ \mu} = 1 ~~, ~k = 1, 2, 3 ~.
\end{equation}
In calculation of $b_{j}^{\mu}~b_{k~ \mu}$ we will use the fact that it
represents scalar product of two four-vectors. Thus, its value is independent
on the frame of reference. For the proper frame of reference
\begin{equation}\label{14}
b_{j}^{\mu}~b_{k~ \mu} = 1 ~-~ \vec{e}_{j} ' \cdot \vec{e}_{k} ' =
		      1 ~-~ \delta_{j~k} ~~, ~j = 1, 2, 3 ~~, ~k = 1, 2, 3 ~,
\end{equation}
since in the optics of scattering processes it is supposed
(defined) that unit vectors
$\vec{e}_{1} ' \equiv \vec{S} '$, $\vec{e}_{2} '$ and $\vec{e}_{3} '$
are orthogonal. Thus, we obtain (inserting results of Eqs. (13) and (14) into
Eq. (12))
\begin{equation}\label{15}
\left ( \frac{d~ p^{\mu}}{d~ \tau} \right ) _{I} ~b_{k~ \mu}  = -~
		   \frac{w^{2}~S~A'}{c} ~Q'_{k} ~, ~~ k = 1, 2, 3 ~.
\end{equation}

Multiplication of Eq. (11) by four-vector $b_{k~\mu}$ (and summation over
$\mu$) yields
\begin{equation}\label{16}
\left ( \frac{d~ p^{\mu}}{d~ \tau} \right ) _{II} ~b_{k~ \mu}  =
		   \frac{w^{2}~S}{c} ~G^{\mu~\nu} ~ b_{\nu} ~b_{k~ \mu}
				~, ~~k = 1, 2, 3 ~.
\end{equation}
The value value is independent on the frame of reference.
For the proper frame of reference the relations of Eq. (5) yield
\begin{eqnarray}\label{17}
\left ( \frac{d~ p^{\mu}}{d~ \tau} \right ) _{II} ~b_{k~ \mu}  &=& ~-
		   \frac{w^{2}~S}{c} ~
	 C'_{~j~i} ~  (\vec{S} ')_{i} ~ ( \vec{e}_{k} ')_{j}
\nonumber \\
&\equiv&  -~ \frac{w^{2}~S}{c} ~
	 ( \vec{e}_{k} ')^{T} ~ ( C'~ \vec{S} ') ~, ~~ k = 1, 2, 3 ~.
\end{eqnarray}
It was already shown (see Eqs. (6) and (7) in Kla\v{c}ka (2001))
that right-hand sides of Eqs. (15) and (17) are identical.
Thus, also left-hand sides of Eqs. (15) and (17) are identical:
\begin{equation}\label{18}
\left ( \frac{d~ p^{\mu}}{d~ \tau} \right ) _{I} ~b_{k~ \mu} =
\left ( \frac{d~ p^{\mu}}{d~ \tau} \right ) _{II} ~b_{k~ \mu} ~, ~~ k = 1, 2, 3 ~.
\end{equation}
If we take into account that four-vectors $b_{k}^{\mu}$
may be taken in various ways, Eq. (18) yields
\begin{equation}\label{19}
\left ( \frac{d~ p^{\mu}}{d~ \tau} \right ) _{I} =
\left ( \frac{d~ p^{\mu}}{d~ \tau} \right ) _{II}  ~.
\end{equation}
Thus, Eqs. (10) and (11) are equivalent, q. e. d..

\section{Conclusion}

We have shown that equation of motion for real, arbitrarily
shaped particle under the action of electromagnetic radiation (parallel
beam of photons) can be written in the form of Eq. (3).
Moreover, it was shown that this equation of motion is equivalent to
a more simple/transparent form for practical calculations --
to Eq. (28) presented in Kla\v{c}ka (2000a).

\vspace*{0.5cm}

{\bf Acknowledgements:} This work was supported by the VEGA grant
No.1/7067/20.


\begin{thebibliography}{}
\bibitem{}Kla\v{c}ka J., 2000a, Electromagnetic radiation and motion of
real particle, http://xxx.lanl.gov/abs/astro-ph/0008510
\bibitem{}Kla\v{c}ka J., 2000b, Aberration of light and motion of real particle,
http://xxx.lanl.gov/abs/astro-ph/0009108
\bibitem{}Kla\v{c}ka J., 2000c, Solar radiation and asteroidal motion,
http://xxx.lanl.gov/abs/astro-ph/0009109
\bibitem{}Kla\v{c}ka J., 2001, Electromagnetic radiation and motion of
arbitrarily shaped particle, http://xxx.lanl.gov/abs/astro-ph/0107123
\bibitem{}Kla\v{c}ka J., Kocifaj M., 2001, Motion of nonspherical dust
particle under the action of electromagnetic radiation,
{\it J. Quant. Spectrosc. Radiat. Transfer} {\bf 70/4-6}, 595-610
\end{thebibliography}
\end{document}